\documentclass[10pt]{article}
\usepackage[latin1]{inputenc}
\usepackage{multicol}
\usepackage{amsfonts}
\usepackage{amssymb}
\usepackage{amsmath}
\begin{document}
\title{Protocol Channels}
\author{Steffen Wendzel\thanks{\texttt{swendzel (at) ploetner-it (dot) de, www.wendzel.de}}\\Kempten University of Applied Science}
\date{2009-07-26}
\maketitle
\begin{abstract}
Covert channel techniques are used by attackers to transfer data in a way prohibited by the security policy. There are two main categories of covert channels: timing channels and storage channels. This paper introduces a new storage channel technique called a \emph{protocol channel}. A protocol channel switches one of at least two protocols to send a bit combination to a destination. The main goal of a protocol channel is that packets containing covert information look equal to all other packets within a network, what makes a protocol channel hard to detect.
\end{abstract}


\begin{small}


\section*{Protocol Channels}

For attackers, it is usual to transfer different kinds of hidden information trough hacked or public networks. The solution for this task can be to use a network covert channel technique like they are well known since years. There are currently two different types of covert channels, so called \emph{storage channels} (which include hidden information in attributes of transfered network packets) and \emph{timing channels} (which make use of the timings of sent packets to transfer hidden information) [Owens02].

A new storage channel technique called a ``protocol channel'' includes hidden information only in the header part of protocols that specify an embedded protocol (e.g. the field ``Ether Type'' in Ethernet, the ``Protocol'' value in PPP, the ``Next Header'' value in IPv6 or the source/destination port of TCP and UDP). For instance, if a protocol channel would use the two protocols ICMP and ARP, while ICMP means that a 0 bit was transfered and ARP means that a 1 bit was transfered, then the packet combination sent to transfer the bit combination ``0011'' would be ICMP, ICMP, ARP, ARP. A protocol channel must not contain any other information that identifies the channel. It is also important that a protocol channel only uses usual protocols of the given network. An algorithm to identify such usual protocols for adaptive covert channels (protocol hopping covert channels) was introduced by [YADALI08].

The higher the number of available protocols for a protocol channel, the higher amount of information can be transfered within one packet since more states are available. Given the above example, two different states are available, what represents 1 bit per packet. If the attacker could use four different protocols, a packet would represent two bits.

Short bit combinations do not allow high covert channel transfer rates but are enough to transfer sniffed passwords or other tiny information. Specially if the attacker uses some compressing algorithm (like converting 7 bit ASCII input to a 6 bit representation of the most important printable characters), the need for a high transfer rate decreases. The proof of concept code ``pct'' uses a minimalized 5 bit ASCII encoding and a 6th bit as a parity bit.

\section*{Problems}

Since a protocol channel only contains one or two (usually not more) bits of hidden information per packet, it is not possibly to include reliability information (like an ACK flag or a sequence number). If a normal packet, that does not belong to the protocol channel, would be accepted by the receiver of a protocol channel, the whole channel would become desyncronized. It is not possibly to identify packets which (not) belong to the protocol channel if they use one of the protocols exploited by the protocol channel.

Another problem is the defragmentation as well as the loss of packets. If a packet is getting fragmented, the receiver receives it two times what means that the bit combination would be used twice and the receiver-side bit combination would be destroyed. The channel would end up desyncronized in this case too. The receiver could check for packets that include the ``More Fragments'' flag of IPv4 as a solution for this problem. Lost packets create a hole in the bit combination what results in the same desyncronization problem.

\section*{Conclusion}
Protocol channels provide attackers a new way to send hidden information through networks. Even if a detection by network security monitoring systems is possible -- e.g. because of unusual protocols used by the attacker -- a regeneration of the hidden data is as good as impossible since it would need information about the transfered data type, the way the sent protocol combinations are interpreted (e.g. big-endian or little-endian) and a recording of all sent packets to enable a regeneration of a channel's input.

\end{small}

\end{document}